\documentclass[a4paper,12pt]{article}
\usepackage{epsfig}
\usepackage{graphicx}
\usepackage{dcolumn}
\usepackage{bm}
\usepackage{longtable}
\usepackage{amssymb}
\usepackage{amsfonts}
\usepackage[margin=1.0in]{geometry}
\usepackage{amsmath}
\usepackage{slashed}

\def\Re{\mathop{\rm Re}\nolimits}
\def\Im{\mathop{\rm Im}\nolimits}

\def\IC{\relax\,\hbox{$\inbar\kern-.3em{\rm C}$}}

\def\e{\epsilon}

\DeclareFontFamily{U}{rsf}{}
\DeclareFontShape{U}{rsf}{m}{n}{
  <5> <6> rsfs5 <7> <8> <9> rsfs7 <10-> rsfs10}{}
\DeclareMathAlphabet\Scr{U}{rsf}{m}{n}

\makeatletter                             %%
\@addtoreset{equation}{section}                   %%
\makeatother                              %%
\usepackage{hyperref}

\begin{document}

\begin{flushright}
CERN PH-TH-2015-013
\end{flushright}
\vskip 1cm
\begin{center}
{\bf\LARGE Observations on the Partial Breaking of $N=2$ Rigid Supersymmetry} \\
\vskip 2 cm
{\bf  Laura Andrianopoli$^{a,b}$,  Riccardo D'Auria$^{a,b}$, Sergio Ferrara$^{c,d}$\footnote{On leave of absence from Department of Physics and Astronomy, U.C.L.A., Los Angeles CA USA} and  \\
Mario Trigiante$^{a,b}$}
\vskip 8mm
 \end{center}
\noindent {$^a$\small{\it DISAT, Politecnico di Torino, Corso Duca
    degli Abruzzi 24, I-10129 Turin, Italy}\\
    \noindent{$^b$ Istituto Nazionale di
    Fisica Nucleare (INFN) Sezione di Torino, Italy}

\noindent {$^c$\small{\it Department of Physics, CERN Theory Division, CH - 1211 Geneva 23, Switzerland }\\
\noindent{$^d$Istituto Nazionale di
 Fisica Nucleare (INFN) Laboratori Nazionali di Frascati, Italy}\\

\vskip 1 cm

\begin{abstract}
We study the partial breaking of $N=2$ rigid supersymmetry for a generic rigid special geometry of $n$
abelian vector multiplets in the presence of Fayet-Iliopoulos terms induced by the Hyper-K\"ahler momentum map.
By exhibiting the symplectic structure of the problem we give invariant conditions for the breaking to occur,
which rely on a quartic invariant of the Fayet-Iliopoulos charges as well as on a modification of the $N=2$ rigid symmetry algebra by a vector central charge.

\end{abstract}
\section{Introduction}
It is well known that partial breaking of rigid and local supersymmetry can occur \cite{Ferrara:1985gj,Cecotti:1985mx}, provided one evades \cite{Cecotti:1985sf,Cecotti:1984wn,Hughes:1986dn,Hughes:1986fa,Ferrara:1995gu,Antoniadis:1995vb,Partouche:1996yp,Ferrara:1995xi,Fre:1996js} some no-go theorems \cite{Witten:1982df,Cecotti:1984rk,Cecotti:1985mx} which are satisfied by a certain class of theories.
%,Cecotti:1984wn,Cecotti:1985sf,,
In global supersymmetry, Hughes and Polchinski first pointed out the possibility  to realize partial breaking of global supersymmetry \cite{Hughes:1986dn} and, in four dimensional gauge theories, this was realized for a model of a self-interacting $N=2$ vector multiplet, in the presence of $N=2$ electric and magnetic Fayet-Iliopoulos terms \cite{Antoniadis:1995vb}. This model is closely connected to the Goldstone action of partially broken $N=2$ supersymmetry \cite{Bagger:1996wp} by integrating out the ($N=1$) chiral-multiplets components of the $N=2$ vector multiplet \cite{Rocek:1997hi}, thus reproducing the supersymmetric Born-Infeld action \cite{Deser:1980ck,Cecotti:1986gb}.
Multi-field versions which generalize the supersymmetric Born-Infeld theory to an  arbitrary number of vector multiplets were then obtained, preserving $N=1$ supersymmetry \cite{Rocek:1997hi,Aschieri:1999jr}, or preserving a second non-linearly realized supersymmetry \cite{Ferrara:2014oka,Andrianopoli:2014mia}.

It is the aim of the present note to further elucidate some general conditions for  partial supersymmetry breaking to occur which are independent on the particular alignment of the unbroken supersymmetry with respect to the original two supersymmetries, and are also independent of the particular representative of the Fayet-Iliopoulos charge vector which, in our problem, is a triplet of the $N=2$ $SU(2)$ R-symmetry and a symplectic vector with respect to the symplectic structure of the underlying Special Geometry: $\mathcal{P}^{M\,x}= \sqrt 2\begin{pmatrix}m^{I\,x}\\e_I^x\end{pmatrix}$. In terms of it, the Ward identities are manifestly invariant on the choice of the symplectic frame. The symplectic invariance relies on the existence of a quartic invariant which is the squared norm of  the $SU(2)$ triplet of symplectic singlets:
\begin{equation}
\xi_x= \frac 12 \epsilon_{xyz}\mathcal{P}^{yM}\mathcal{P}^{zN}\mathbb{C}_{MN}= 2 \left( \vec m^I\times\vec e_I\right)_x\,.
\end{equation}

We shall give, in Sect. 2, the general Ward identities that the $N=2$ scalar potential satisfies when Fayet-Iliopoulos symplectic-charge triplets $\mathcal{P}^{xM}$ are turned on, explicitly showing that they are modified by a constant traceless matrix $C^A{}_B= \xi_x (\sigma^x)^A{}_B$.
Furthermore, in Sect. 2.1,  we shall derive in a symplectic covariant manner the modifications of the supersymmetry algebra which, in the framework of $N=2$ tensor calculus, was derived in \cite{Ferrara:1995xi}.

\section{The rigid Ward Identity}
It is a well known fact that the Supergravity Ward identity relating the scalar potential $V$ to the  shifts of the fermions in the presence of a gauging is a pure trace identity in the R-symmetry ${\rm SU}(N)$ indices, namely:

\begin{equation}\label{ward}
    V\,\delta^A_B =
    \sum_i\alpha_i\,  \delta\chi^{i A}\delta \chi_{iB}
\end{equation}
where the index $i$ in the sum runs over all the fermion-shifts of the theory (including the gravitino), $\alpha_i$ being constants which are positive for the spin $1/2$--fields and negative for the gravitino. This is true also when  the
scalar potential is due to the presence of a Fayet-Iliopoulos (FI) term \cite{Ferrara:1985gj,Cecotti:1985mx,Cecotti:1984wn,Andrianopoli:1996cm}.

In the rigid supersymmetric theories with $N>1$ the previous statement is violated, since on the right-hand-side of equation (\ref{ward}) a traceless term  can appear related to the presence of electric and \emph{magnetic} FI terms \cite{Hughes:1986dn,Antoniadis:1995vb,Ferrara:1995xi,Rocek:1997hi}. In the case of a $N=2$ rigid theory, this can be seen either by direct computation of the fermionic shifts of the gauginos or by performing a suitable flat limit of the $N=2$ Supergravity  parent theory with gravitino and hyperinos constant non zero shifts.

 In the rigid case  of a supersymmetric vector-multiplet theory, equation (\ref{ward}) allows for a traceless constant term $C^A_B$ in the scalar potential Ward identity, namely (\ref{ward}) is modified as follows

\begin{equation}\label{modif}
    V\,\delta^A_B + C_B{}^A= \sum_i \delta \lambda^{i A}\delta \lambda_{iB}.
\end{equation}
where $\lambda^{iA}$ and $\lambda_{iA}\equiv g_{i\bar k}\lambda^{\bar k}_A$ denote the chiral and antichiral projections of the gauginos, respectively. According to the arguments given in \cite{Ferrara:1995xi}, the additive constant matrix $C_B{}^A$ can be interpreted as a central extension in the supersymmetry algebra, which only affects the commutator of two supersymmetry transformations of the gauge field.

In the $N=2$ case the traceless matrix $C^A_{\,\,B}$ has the following expression
\begin{equation}\label{cab}
 C_B{}^A = \frac12\epsilon_{xyz}\mathcal{P}^{yM}\mathbb{C}_{MN}\mathcal{P}^{zN}\, (\sigma^{x})_B{}^A\equiv \xi^x\, (\sigma^{x})_B{}^A
\end{equation}
where $(M,N,...)$ are symplectic indices in the fundamental of ${\rm Sp}(2n)$,

\begin{equation}\label{pp}
   \mathcal{P}^{x}_M=\sqrt 2\left(
           \begin{array}{c}
             -e_{I}^x \\
             m^{I\,x}\\
           \end{array}
         \right)= -\mathbb{C}_{MN}\,\mathcal{P}^{N\,x}
\end{equation}
is a constant symplectic vector (defining the electric/magnetic Fayet-Iliopoulos term) whose upper and lower components $I=(1,2,...n)$ are electric and magnetic respectively, and

 \begin{equation}
\mathbb{C}_{MN}=
\left(
  \begin{array}{cc}
    \textbf{0} & \textbf{1} \\
    \textbf{-1} & \textbf{0} \\
  \end{array}
\right)
\end{equation}
is the symplectic metric.

Let us give for completeness the derivation of this result in the case of $N=2$ supersymmetric theory with a number $n$ of abelian vector multiplets, no hypermultiplets and in the presence of both electric and magnetic Fayet-Iliopoulos terms.

In this case, the fermion-shift in the supersymmetry transformation of the chiral gaugino fields can be written, using a symplectic formalism, as
\begin{equation}\label{funda}
 \delta \lambda^{iA}=W^{iAB}\epsilon_B
\end{equation}
with
 \begin{equation}\label{W}
   W^{iAB}= i\,(\sigma^{x})_C{}^A\epsilon^{CB}\,\mathcal{P}^{x}_M\,g^{i\bar k}\bar U^M_{\bar k}
 \end{equation}
 where $g_{i\bar k}$ is the rigid Special K\"ahler metric and the symplectic section $U^M_{i}$ is the derivative with respect to the scalar fields $z^i$ of the fundamental symplectic section $V^M(z)$ of the rigid Special Kaehler Geometry \cite{Seiberg:1994rs,Ceresole:1995jg}:

 \begin{equation}\label{Ui}
   U^M_{i}=\frac{\partial}{\partial z^i}\left(
                                          \begin{array}{c}
                                            X^I \\
                                            F_I\\
                                          \end{array}
                                        \right)=\frac{\partial}{\partial z^i}V^M(z)\,.
 \end{equation}
 Introducing a triplet of triholomorphic superpotentials $\mathcal{W}^x$

\begin{equation}\label{super}
   \mathcal{W}^x=\mathcal{P}^x_M \,V^M(z)=\sqrt 2\left(F_I(z) m^{xI} -e_I^x \,X^I (z)\right)
\end{equation}
equation (\ref{funda}) takes the form

\begin{equation}\label{new}
  \delta \lambda^{iA}= i(\sigma^{x})_C{}^A\epsilon^{CB}\,\partial_{\bar k} {\bar{\mathcal{W}}}^x g^{i\bar  k}\epsilon_B=i\,Y^{ix}(\sigma^{x})_C{}^A\epsilon^{CB} \epsilon_B
\end{equation}
with $Y^{ix}=g^{i\bar  k}\,\partial_{\bar k} {\overline{\mathcal{W}}}^x$.

Let us now use special coordinates, that is $X^I=z^i$; in this frame we can write

\begin{eqnarray}\label{special}
 \nonumber   g_{IJ}&=&\Im { F}_{IJ}\\
    U^M_I&=&\left(
            \begin{array}{c}
              \delta_I^K \\
              F_{IK} \\
            \end{array}
          \right) \longrightarrow g^{I\bar J}\bar U^M_{\bar J}=\left(
                                                                 \begin{array}{c}
                                                                   g^{I\bar K} \\
                                                                   g^{I\bar J} \Re F_{ K J }-i \delta^{ I}_{K}\\
                                                                 \end{array}
                                                               \right)
\end{eqnarray}
 A short computation then gives

 \begin{equation}\label{short}
   Y^{Ix}=\sqrt 2\,\left[-g^{I\bar K}\left( e^x_K-\Re F_{KJ}\,m^{Kx}\right)-im^{Ix}\right]\,.
 \end{equation}
 This formula actually coincides with eq. (23) of \cite{Antoniadis:1995vb} and it shows that a non-zero magnetic charge $m^{Ix}$ produces a constant imaginary part of the auxiliary field $ Y^{Ix}$, a necessary condition for partial breaking of supersymmetry.

We now use the Special Geometry identity \cite{Andrianopoli:2006ub}:
\begin{equation}\label{identity}
    U^{MN}=U^M_i g^{i\bar k}\,\bar U^N_{\bar k}=\frac12\left( \mathcal M^{MN}-i \mathbb{C}^{MN}\right)\,,
\end{equation}
where $\mathcal{M}^{MP} \mathcal{M}_{PN}=\delta^M_N$ and 
\begin{eqnarray}
{\mathcal M}_{MN}=-\begin{pmatrix}
{I}+ {R}\cdot {  I}^{-1} \cdot {R} & -{ R}\cdot { I}^{-1}\\
-{  I}^{-1}\cdot  { R} & {  I}^{-1}
\end{pmatrix}>0\,,\label{Mgt2}
\end{eqnarray}
$I\equiv (-{\rm Im}(F_{IJ})$ and $R\equiv ({\rm Re}(F_{IJ})$. 
If we  flatten the $\sigma$--model coordinate index of $\delta \lambda^{iA}$ in (\ref{funda}), we obtain
\begin{equation}\label{flatten}
U^N_i\,\delta \lambda^{iA}=\frac{i}{2}\left( \mathcal M^{MN}-i \mathbb{C}^{MN}\right)\mathcal{P}_N^x\,(\sigma^{x})_C{}^A\epsilon^{CB}\epsilon_B\,.
\end{equation}
Finally we may compute the bilinear product in the gaugino shifts
\begin{equation}\label{fin}
   g_{i\bar k}W^{iAB}\bar W^{\bar k}_{BC}
      =\frac{\delta_C^A}{2} \mathcal M^{MN}\,\mathcal{P}^x_M \,\mathcal{P}^x_N + \frac12 \mathbb{C}^{MN}\,\mathcal{P}^x_M \,\mathcal{P}^x_N \epsilon^{xyz}(\sigma_z)_C{}^A= \delta^A_C\,V_{N=2}+ C_C{}^A\,,
    \end{equation}
which coincides with equation (\ref{modif}), proving our statement.
In conclusion, the $N=2$ scalar potential of the rigid theory is
\begin{equation}\label{concl}
 V_{N=2}= \frac{1}{2}(\mathcal{P}^x)^T\,{\mathcal{M}^{-1}}\,\mathcal{P}^x
\end{equation}
while, by the identification (\ref{pp}),  the $C^A_{\,\,B}$ term can be rewritten as

\begin{equation}\label{rewri}
   C_A{}^B = \frac12\, (\sigma^{x})_A{}^B\,\mathbb{C}^{MN}\,\mathcal{P}^y_M \,\mathcal{P}^z_N \epsilon_{xyz}=2\,\left(\vec{m}^I\times \vec{e}_I\right)^x\, (\sigma^{x})_A{}^B\,.
\end{equation}

In the general case in which both the F and D-terms are present, we define the following ${\rm SO}(3)$-vector:
\begin{equation}
\xi^x\equiv \frac{1}{2}\,\epsilon_{xyz}\,\mathcal{\mathcal{P}}^y_M\,\mathcal{\mathcal{P}}^z_N\mathbb{C}^{MN}=2\,(\vec{m}^I\times \vec{e}_I)^x\,,
\end{equation}
where $\vec{m}^I\equiv (m^{I\,x})\,\,,\,\,\vec{e}_I\equiv (e_{I}{}^{x})$. In terms of this quantity
eq. (\ref{fin}) reads:
\begin{equation}
  g_{i\bar k}W^{iAB}\bar W^{\bar k}_{BC}
     = \delta^A_C\,V_{N=2}+ C_C{}^A=\left(\begin{matrix}V_{N=2}+\xi^3 & \xi^1-i\,\xi^2\cr  \xi^1+i\,\xi^2 & V_{N=2}-\xi^3\end{matrix}\right)\,.\label{deldel}
\end{equation}
Upon diagonalization, the above matrix reads
\begin{equation}
 g_{i\bar k}W^{iAB}\bar W^{\bar k}_{BC}
     = \left(\begin{matrix}V_{N=2}+\sqrt{|\xi|^2} &0\cr  0 & V_{N=2}-\sqrt{|\xi|^2}\end{matrix}\right)\,,\label{diagbasis}
\end{equation}
where $|\xi|^2$ denotes the following quartic invariant $I_4$ in the FI terms:
\begin{equation}
I_4\equiv |\xi|^2= \xi^x\,\xi^x=\frac{1}{2}\sum_{x,y}(\mathcal{\mathcal{P}}_M^x\mathcal{\mathcal{P}}_N^y\,\mathbb{C}^{MN})^2=4\,(\vec{m}^I\cdot \vec{m}^J \,\vec{e}_I\cdot \vec{e}_J -\vec{m}^I\cdot \vec{e}_J \,\vec{m}^J \cdot \vec{e}_I)\,.
\end{equation}
From eq. (\ref{deldel}) we observe that the square root of the quartic invariant $|\xi|^2$ defines the fourth power of the  supersymmetry breaking scale.\par
Let us discuss the supersymmetry breaking patterns.
\begin{itemize}
\item{If $I_4\neq 0$, $N=2$ is spontaneously broken to either $N=1$ or $N=0$, depending on whether $V_{N=2}>\sqrt{I_4}$ or  $V_{N=2}=\sqrt{I_4}$, respectively. In the latter case one of the eigenvalues vanishes and the corresponding direction in superspace defines the surviving $N=1$ supersymmetry. The $N=1$ potential is the square of the fermion shifts along the
    direction of the residual supersymmetry. In the diagonal basis (\ref{diagbasis}), this direction is the second one so that:
    \begin{equation}
V_{N=1}=g_{i\bar k}W^{i22}\bar W^{\bar k}_{22}=V_{N=2}-\sqrt{|\xi|^2}\,.
    \end{equation}
In this case, the infra-red dynamics is captured by a Born-Infeld Lagrangian \cite{Ferrara:2014oka}.
    }
    \item{If $I_4= 0$, which implies  $\xi^x=2\,(\vec{m}^I\times \vec{e}_I)^x=0$,
     when $V_{N=2}\neq 0$  $N=2$ is completely broken.
Supersymmetry can be preserved only at the boundary of the moduli space if $\mathcal{P}_M^x\neq 0$, or everywhere if $\mathcal{P}_M^x=0$ (in which case there is no potential).}
\end{itemize}

In the absence of D-terms, $\xi^1=\xi^2=0$ and $\sqrt{|\xi|^2}=|\xi^3|$. Taking for instance $\vec{m}^I=(m^I,0,0)$ and $\vec{e}_I=(0,e_I,0)$,
we find
\begin{equation}
\xi^1=\xi^2=0\,\,,\,\,\,\,\xi^3=2\,m^I\,e_I\,.
\end{equation}
In this case eq. (\ref{deldel}) becomes:
\begin{equation}
  g_{i\bar k}W^{iAB}\bar W^{\bar k}_{BC}=
  \left(\begin{matrix}V_{N=2}+\xi^3 & 0\cr  0& V_{N=2}-\xi^3\end{matrix}\right)=  \left(
            \begin{matrix}
             \bar P(\mathcal {M}-i\mathbb{C})\,P&0 \cr
             0& \bar P(\mathcal {M}+i\mathbb{C})\,P
            \end{matrix}\right)\,,\label{deldel2}
\end{equation}
where we have defined $P^M=\frac{1}{\sqrt 2} {\mathbb{C}}^{MN} \left(\mathcal{P}^1_N+i\mathcal{P}^2_N\right)$. If %$0<2\,m^I\,e_I=\xi^3=\sqrt{|\xi|^2}$,
$\xi^3>0$, the residual $N=1$ supersymmetry is along the second direction ($\epsilon_2$), while, if $\xi^3<0$, along the first. In the former case the lower diagonal entry of (\ref{deldel2}), along the direction of the preserved supersymmetry, defines the $N=1$ potential:
\begin{equation}
V_{N=1}=V_{N=2}-\xi^3= P(\mathcal {M}+i\mathbb{C})\,P=\bar{P}\mathcal{M}P-2\,m^I\,e_I\,.
\end{equation}
This is consistent with (\ref{flatten}) which, in the absence of a D-term, can be written in the form
\begin{equation}\label{flatten2}
  \mathbb{C}_{MN}\,U^N_i\,\delta \lambda^{iA}=\frac{i}{\sqrt{2}}\left(\begin{matrix}-(\mathcal{M}+i\,\mathbb{C})_{MN}P^N\,\epsilon_1\cr (\mathcal{M}-i\,\mathbb{C})_{MN}\bar{P}^N\,\epsilon_2\end{matrix}\right)\,.
\end{equation}
Indeed at the $N=1$ vacuum
\begin{equation}
 \mathbb{C}_{MN}\,U^N_i\,\delta_2 \lambda^{iA}=\frac{i}{\sqrt{2}}\,(\mathcal{M}-i\,\mathbb{C})_{MN}\bar{P}^N\,\epsilon_2=0\,\,\Rightarrow\,\,\,V_{N=1}=0\,.
\end{equation}
\subsection{Supersymmetry Transformation of the Vector Fields}
Let us introduce, besides the electric vector fields $A^I_\mu$, the magnetic ones $A_{I\,\mu}$. The supersymemtry transformation property of the former can be extended to the latter in a symplectic covariant fashion. Define the symplectic vector:
\begin{equation}
A^M_\mu\equiv \left(\begin{matrix}A^I_\mu\cr A_{I\,\mu}\end{matrix}\right)\,.
\end{equation}
The supersymmetry transformation property of $A^M_\mu$ reads:
\begin{equation}
\delta A^M_\mu=i\,U_i^M\,\bar{\lambda}^{iA}\gamma_\mu \epsilon^B\epsilon_{AB}+h.c.
\end{equation}
Using eq.s (\ref{funda}) and (\ref{W}) we can evaluate the commutator of two supersymmetry transformations  on $A^M_\mu$:
\begin{align}
\delta_{[1}\delta_{2]}\,A^M_\mu &=-U_i^M\,g^{i\bar{\jmath}}\,\overline{U}_{\bar{\jmath}}^N (\sigma^{x})_C{}^A\epsilon^{CB}
\bar{\epsilon}_{[2\,B} \gamma_\mu\,\epsilon_{1]}^D\epsilon_{AD}\,\mathcal{P}^x_N+h.c.=\nonumber\\&=
(U^{MN}-\overline{U}^{MN})\,\mathcal{P}^x_N\,(\sigma^{x})_A{}^B\,\bar{\epsilon}_{[2\,B} \gamma_\mu\,\epsilon_{1]}^A=-i\,\mathbb{C}^{MN}\,\mathcal{P}^x_N\,(\sigma^{x})_A{}^B\,\bar{\epsilon}_{[2\,B} \gamma_\mu\,\epsilon_{1]}^A\,,
\end{align}
where we have used (\ref{identity}).
From the above expression we conclude that:
\begin{equation}
\delta A^I_\mu=-i\,\sqrt{2}\,m^{I\,x}(\sigma^{x})_A{}^B\,\bar{\epsilon}_{[2\,B}\gamma_\mu\,\epsilon_{1]}^A\,,
\end{equation}
that is the commutator of two supersymmetry transformations on a vector field, in the presence of a magnetic Fayet-Iliopoulos term, amounts to a shift, whose parameter corresponds to a  vector central charge in the $SU(2)$ adjoint representation \cite{Ferrara:1995xi,Dvali:1996xe,Ferrara:1997tx}.
\section*{Acknowledgements}
S.F. would like to thank G. Dvali, A.Marrani, M.Porrati and A.Sagnotti for enlightening discussions. S.F. would like also to thank CERN for its kind hospitality.

\end{document}